\documentstyle[preprint,aps,prd]{revtex}

\begin{document}

\title {Exact  CTP Renormalization Group Equation for the Coarse
Grained Effective Action}
    
\author{Diego A.\ R.\ Dalvit \footnote{Electronic address: dalvit@df.uba.ar}}

\address{{\it
Departamento de F\'\i sica, Facultad de Ciencias Exactas y Naturales\\ 
Universidad de Buenos Aires- Ciudad Universitaria, Pabell\' on I\\ 
1428 Buenos Aires, Argentina}}

\author{Francisco D.\ Mazzitelli 
\footnote{Electronic address: fmazzi@df.uba.ar}}

\address{{\it
Departamento de F\'\i sica, Facultad de Ciencias Exactas y Naturales\\ 
Universidad de Buenos Aires- Ciudad Universitaria, Pabell\' on I\\ 
1428 Buenos Aires, Argentina\\
and\\
Instituto de Astronom\'\i a y F\'\i sica del Espacio\\
 Casilla de Correo 67 - Sucursal 28\\
1428 Buenos Aires, Argentina}}

\maketitle

\begin{abstract}
We consider a scalar field theory in Minkowski spacetime
and define a coarse grained Closed Time Path (CTP) effective action 
by integrating quantum fluctuations of wavelengths
shorter than a critical value. We derive an exact CTP   
renormalization group equation 
for the dependence of the effective action on the coarse graining scale.
We solve this equation using a derivative expansion 
approach. Explicit calculation
is performed for the $\lambda \phi^4$ theory. We discuss the 
relevance of the CTP average action in the study of non-equilibrium
aspects of phase transitions in quantum field theory. 
\end{abstract}
\newpage

\section{INTRODUCTION}

The study of phase transitions in Quantum Field Theory is of
great interest in cosmology and particle physics. Cosmological
inflationary
models \cite{kolb,mazenko,boya1}, 
the electroweak phase transition \cite{bastian,kolbglei}, the formation of
chiral condensates \cite{raja,jpp1,boya2}, are 
clear examples of 
the very
interesting problems related to phase transitions in this
context.

The first analyses of phase transitions in Quantum Field Theory
were based on the use of the finite temperature effective
potential. The effective potential
is useful only in quasi-static situations and therefore one must
use the complete effective action in order to address
the non-equilibrium aspects of the transition.     

The  usual effective action is not adequate to 
study  initial value problems since it gives the evolution
equations for
`in-out'  matrix elements of the background fields. 
The equations for
these matrix elements are neither real nor causal. The solution
to this problem is to use the so called `in-in' or Closed
Time Path (CTP) effective action, introduced by Schwinger
and Keldysh many years ago \cite{ctp}. 
This action gives real and causal evolution equations
for the `in-in' mean value of the background fields.
The CTP effective action has been used to analyze
inflationary models,
anisotropy dissipation in the early universe, 
the backreaction
problem in semiclassical  and stochastic gravity,
the quantum to classical transition in quantum Brownian motion
and Quantum Field Theory, etc. \cite{hu1,hu2}.

There is however one aspect that, to our knowledge, has not
been fully investigated.  Phase transitions occur via the
formation and growth of {\it spatial} domains. Inside these domains,
the order parameter  of the transition evolves dynamically,
and one is usually interested in computing its temporal
evolution. 
The order parameter is therefore {\it the average of the
quantum mean value of the field in a spatial volume
of the size of the domain}.

In previous works, this problem has been addressed using
different approaches. On the one hand, phase
transitions have been analyzed using the CTP effective action,
assuming that the order parameter depends only on time.
Presumably, this time dependent function describes the 
dynamics of the order parameter inside one typical domain \cite{bdv}.
It has been shown that domain growth is an effect characterized
by the rapid evolution of (exponentially unstable) long wavelength
modes. Such a dynamics can be non-perturbatively described by a
Hartree approximation to the two-point correlation function
\cite{hartree}.
On the other hand,
phenomenological Langevin-like equations which account for 
dissipation and noise have been proposed 
and  numerically solved 
(see Ref. \cite{gleiser} for one such type of calculation, 
where the order parameter is coupled to a thermal bath and a 
Markovian Langevin equation is put forward in order to mimic thermal 
phase mixing during a first-order phase transition).

In this paper, as a step towards a first principle calculation, 
we will proceed using an
analogy with what is done in the context of 
condensed  matter \cite{goldenfeld}.   
We will coarse grain our theory up to
a length scale $\Lambda^{-1}$ comparable to the initial size of a
typical domain.
In this way, we will define a `Coarse Grained Effective Action'
(CGEA),
which will be basically the usual CTP action in which only those
modes of the scalar field with $\vert\vec q\vert > \Lambda$
are integrated out. As a result of tracing the short wavelength 
modes (the so-called ``environmental" degrees of freedom), 
the CTP average action, which depends on the long wavelength 
modes (the so-called ``system" degrees of freedom), develops 
dissipation and noise terms. 

We will derive an exact evolution equation for the dependence of
the CGEA on the coarse graining scale.
In principle it is possible to derive a Langevin  
equation for the long wavelength order parameter 
$\phi({\vec x},t)$ starting from
the CGEA.  
However, in order to make the analysis more tractable, 
here we will compute the CTP average action using some 
simple appro\-xi\-mations. 
The major approximation we will make is to ignore spatial correlations 
between different domains. This allows the study of the 
dynamics for coarse grained time-dependent configurations 
$\phi(t)$ inside a given domain.

The paper is organized as follows: in Section 2 we define the CGEA 
and compute it
in the one loop approximation, making an adiabatic expansion. 
We should mention that by using such approximations we are not
aiming at studying domain growth since, as we have already stated,
they fail to describe such a dynamical process. Rather our intention
is to illustrate a simple calculation of the CTP effective action. 
In Section 3 we derive the exact Renormalization Group (RG) equation 
for the evolution of the CGEA,  which is a generalization to 
CTP of the euclidean Wegner and 
Houghton's RG equation \cite{wegner}. Using a derivative expansion,
in Section IV we reduce this functional equation to a system of
coupled differential equations, which are then numerically 
solved. The interest of the results is not only restricted to non-zero 
coarse graining scales, but they also provide non perturbative 
approximations to standard ($\Lambda=0$) quantum field theory. 
In Section 5 we make the conclusions.

In order to make contact with related works, we would like to
mention that the CGEA defined here was originally introduced in Ref.
\cite{silarg} in order to study inflationary cosmology, and it was
perturbatively evaluated in Ref. \cite{fernando} in order to analyze the
decoherence of the long wavelength sector of the $\lambda \phi^4$
field theory.
It is also similar 
in spirit to
the {\it euclidean} average effective action proposed 
earlier \cite{wette1,liao1,polsch,hasen,morris1,boni}.
The main difference between both actions
is that the euclidean action averages the field
over a space-time volume, while our CTP action averages
the field over a spatial volume, and it is therefore more
useful to study non-equilibrium situations. 

In the above
mentioned works, and also in this paper, 
the coarse grained or average effective
action interpolates between the bare theory at $\Lambda=\Lambda_0$
(the ultraviolet cutoff) and the physical theory at the
coarse graining scale. Another possibility has been recently
analyzed in Ref. \cite{atta}, where an effective action 
is defined that interpolates between the physical theory
at $T=0$ and the physical theory at $T\neq 0$.  


\section {THE CTP COARSE GRAINED EFFECTIVE ACTION}

Let us consider $\lambda \phi^4$ field theory in Minkowski spacetime. 
In order to deal with non-equilibrium scenarios we 
follow Schwinger-Keldysh formulation, doubling the number of fields and 
imposing CTP boundary conditions. The coarse grained CTP effective action 
$S_{\Lambda}(\phi_{+},\phi_{-})$ is defined as follows
\begin{equation}
e^{i S_{\Lambda}(\phi_+,\phi_-)} \equiv \int \prod_{\Lambda_0>|{\vec q}| > 
\Lambda} {\cal D}[\phi_+({\vec q},t)] 
{\cal D}[\phi_-({\vec q},t)] \; e^{i S_{cl}(\phi_+,\phi_-)}
\end{equation} 
where $S_{cl}(\phi_+,\phi_-) = S_{cl}(\phi_+) - S_{cl}(\phi_-)$. Note that 
a sharp cutoff $\Lambda$ and an ultraviolet cutoff $\Lambda_0$ have been 
used. The functional integrals over the short wavelength modes are to be 
computed using standard CTP boundary conditions: the fields $\phi_+$ 
and $\phi_-$ must have only negative and positive frequency modes,
respectively,
in the past $-T$,  
and match in the future $T$. In general, the CTP coarse grained effective
action has an imaginary part, related to noise, and a real part, related
to dissipation.  The equations of motion derived from it are, however,
real and causal.

An exact calculation of the above defined action is quite difficult, 
and it is therefore necessary to use approximation methods.
One possible approach is to make perturbations in the coupling constant 
$\lambda$ \cite{fernando}.
Another possibility that we will explore here is to use a 
loop expansion. 
In the one loop approximation our effective action can be written as 
$S_{\Lambda}(\phi_+,\phi_-) = S_{cl}(\phi_+) - S_{cl}(\phi_-) 
+ \Delta S_{\Lambda}(\phi_+,\phi_-)$, where the last 
term is linear in $\hbar$. 

As mentioned in the Introduction, we shall be concerned with 
background configurations that depend only 
on time, $\phi_{\pm}(t)$. We split the complete field 
$\phi_{\pm} \rightarrow \phi_{\pm}(t) + \varphi_{\pm}$, where the 
fluctuations $\varphi_{\pm}$ contain spatial modes with $|{\vec q}| > 
\Lambda$. Note that we are assuming that the only mode 
with $|{\vec q}| < \Lambda$ is the spatial homogeneous one 
(${\vec q}=0$). The one loop correction is
\begin{eqnarray}
e^{i \Delta S_{\Lambda}(\phi_+(t),\phi_-(t))}& = &
\int \prod_{\Lambda_0>|{\vec q}| > 
\Lambda} {\cal D}[\varphi_+] {\cal D}[\varphi_-] \;
e^{
\frac{i}{2} \int dt \int \frac{d^3q}{(2\pi)^3} \left[
\varphi_+ \; \frac{\delta^2 S_{cl}}{\delta \phi_+ \delta \phi_+} \; 
\varphi_+ \; - \;
\varphi_- \; \frac{\delta^2 S_{cl}}{\delta \phi_- \delta \phi_-} \;
\varphi_- \right] 
} \times \nonumber \\
&&~~~~~~~~~~~~~~~~~~~~~~~~~~~~~~
e^{
\frac{i}{2} \int^{'} \frac{d^3q}{(2\pi)^3} \int dt
\frac{d}{dt} \left[
\varphi_+ \; \dot{\varphi}_+ \; - \; \varphi_- \; \dot{\varphi}_- 
\right]
}
\end{eqnarray}
where the functional derivatives are evaluated at $\varphi_{\pm}=0$. 
The action for the quantum fluctuations is that of a free field 
with mass $M^2_{\pm}=V''(\phi_{\pm})$, where $V$ is the potential 
in  the (bare) classical action. Their spatial Fourier modes are 
therefore harmonic oscillators with time dependent frequency, 
namely $w^2_{q,{\pm}}(t) = q^2 + V''(\phi_{\pm}(t))$ with 
$q=|{\vec q}|$. The functional integral is quadratic and can be 
done straightforwardly. The result is
\begin{equation}
\Delta S_{\Lambda}(\phi_+(t),\phi_-(t)) =
\frac{i}{2} \int_{\Lambda_{0}>\vert\vec q\vert >\Lambda}  {d^3 q\over (2\pi)^3}
\ln [ g_-({\vec q},T) {\dot g}_+({\vec q},T) - 
      g_+({\vec q},T) {\dot g}_-({\vec q},T) ]
\label{1looP}
\end{equation}
where the modes $g_{\pm}$ are solutions to ${\ddot g_{\pm}}({\vec q},t) + 
w^2_{q,\pm}(t) g_{\pm}({\vec q},t)=0$ satisfying CTP conditions in the 
past and having an arbitrary normalization in the future (we will
present an explicit proof of Eq.(\ref{1looP}) in Section IV).   

Even in the one loop approximation, the effective action
is a very complicated object and additional approximations
are needed in order to get analytic results. 
The simplest approximation one can think of is
the adiabatic approximation \cite{birrel}, in which one
neglects the excitations of the quantum fluctuation field
due to the time dependence of the background field $\phi_{\pm}(t)$.
This approximation misses the very important stochastic
aspects of the theory. However, it will be useful as a warm up
and also to make contact with the euclidean works.

We can write the mode functions as
\begin{equation}
g_{\pm}({\vec q},t) = \frac{1}{\sqrt{2 W_q(\phi_{\pm}(t))}} e^{\pm i 
\int_{-T}^{t}  dt' W_q(\phi_{\pm}(t'))}
\end{equation}
where the functions $W_q(\phi_{\pm})$ satisfy
\begin{equation}
W^2_q + {1\over 2}\left( {\ddot W_q\over W_q}-{3\over 2}\left(
{\dot W_q\over W_q}\right )^2\right)=w^2_q
\label{W}
\end{equation}
The adiabatic approximation consists in solving 
this equation using an expansion
in derivatives of the background field. The result is 
\begin{equation}
W^2_q= q^2 + V''+ {5\over 16}\left[
\left({V'''\over q^2+V''}\right )^2-{V''''\over 4(q^2+V'')}\right ]
\dot\phi^2-{V'''\over 4(q^2+V'')}\ddot \phi + \ldots
\label{Wapprox}
\end{equation}
where the ellipsis denote higher derivative terms.

In the one loop and adiabatic approximations, the average CTP effective
action is therefore, up to a surface term evaluated at $t=T$, given by
$\Delta S_{\Lambda}(\phi_+,\phi_-)=\Delta S_{\Lambda}(\phi_+)
- \Delta S_{\Lambda}(\phi_-)$, where
\begin{equation}
\Delta S_{\Lambda}(\phi)={1\over 2}\int dt\int_{\Lambda}^
{\Lambda_{0}} {d^3 q\over (2\pi)^3}\left (
-\sqrt {q^2+V''}+{\dot\phi^2\over 32}{V'''^2\over (q^2+V'')^{5/2}}
\right )
\label {gamma1loop}
\end{equation}

Including the classical part, we can write the total effective
action as
\begin{equation}
S_{\Lambda}(\phi)=\int dt 
\left ( - V_{\Lambda}(\phi) + {1\over 2}(1+Z_{\Lambda}
(\phi ))\dot\phi^2 + \ldots \right)
\label{s1loop}
\end{equation}
where 
\begin{eqnarray}
V_{\Lambda}&=& V+{1\over 2}\int_{\Lambda}^{\Lambda_{0}}
{d^3 q\over (2\pi)^3}  \sqrt{q^2+V''} = \nonumber\\
&& V + \frac{1}{4 \pi^2} \left[
\frac{\Lambda_0}{4} \sqrt{\Lambda_0^2+V''} (\frac{V''}{2}+\Lambda_0^2) -
\frac{V''^2}{8} \ln(\Lambda_0+\sqrt{\Lambda_0^2+V''}) - \right. \nonumber \\
&& ~~~~~~~~~~~~~~\left. \frac{\Lambda}{4} \sqrt{\Lambda^2+V''} 
(\frac{V''}{2}+\Lambda^2) +
\frac{V''^2}{8} \ln(\Lambda+\sqrt{\Lambda^2+V''}) \right] 
\label{V1loop}
\end{eqnarray}
and
\begin{eqnarray}
Z_{\Lambda}&=&{1\over 32}\int_{\Lambda}^{\Lambda_{0}}
{d^3 q\over (2\pi)^3} {V'''^2\over (q^2+V'')^{5/2}} =
\frac{1}{192 \pi^2} \frac{V'''^2}{V''} \left[
\frac{\Lambda_0^3}{(\Lambda_0^2+V'')^{3/2}} -
\frac{\Lambda^3}{(\Lambda^2+V'')^{3/2}}
\right]  
\label {Z1loop}
\end{eqnarray}
While the function $Z_{\Lambda}(\phi)$ is finite when 
$\Lambda_0 \rightarrow \infty$, the potential $V_{\Lambda}(\phi)$ 
diverges in the UV. In order to make contact with the conventional 
renormalization schemes,we write the one loop bare potential as 
$V(\phi)=\frac{1}{2} (m_R^2+\delta m^2)\phi^2 + \frac{1}{4!} 
(\lambda_R+\delta \lambda) \phi^4$, where the renormalized mass 
and coupling constant are defined as
\begin{eqnarray}
m_R^2 \equiv \left. \frac{\partial^2 V_{\Lambda}}{\partial \phi^2} 
\right|_{\Lambda=\phi=0}
& \; \; \; ; \; \; \; &
\lambda_R \equiv \left. \frac{\partial^4 V_{\Lambda}}{\partial \phi^4} 
\right|_{\Lambda=\phi=0}
\end{eqnarray}
and the counterterms are
\begin{eqnarray}
\delta m^2 &=& -\frac{\lambda_R}{32 \pi^2} \left[
\frac{m_R^2}{2} + 2 \Lambda_0^2 + m_R^2 \ln(\frac{m_R^2}{4 \Lambda_0^2}) 
\right] \nonumber \\
\delta \lambda &=& -\frac{3 \lambda_R^2}{32 \pi^2} \left[
2 + \lambda_R^2 \ln(\frac{m_R^2}{4 \Lambda_0^2})
\right]
\end{eqnarray}
Therefore the renormalized potential is
\begin{eqnarray}
V_{\Lambda}^{ren} (\phi) &=&
\frac{1}{2} m_R^2 \phi^2 (1-\frac{\lambda_R}{64 \pi^2}) +
\frac{1}{4!} \lambda_R \phi^4 (1-\frac{3 \lambda_R}{16 \pi^2}) + \nonumber \\
&& \frac{1}{32 \pi^2} \left[
-\Lambda (2\Lambda^2 + m_R^2 + \frac{1}{2} \lambda_R \phi^2) 
\sqrt{\Lambda^2 + 
m_R^2 +\frac{1}{2} \lambda_R \phi^2} \; + \right. \nonumber \\
&& \left. ~~~~~~~~ (m_R^2+\frac{1}{2} \lambda_R \phi^2)^2 
\ln(\frac{\Lambda+\sqrt{\Lambda^2+m_R^2+\frac{1}{2} \lambda_R \phi^2}}{m_R})
\right]
\end{eqnarray}
and the renormalized wave function renormalization is
\begin{equation}
Z_{\Lambda}^{ren} (\phi) =
\frac{1}{192 \pi^2} \frac{\lambda_R^2 \phi^2}{m_R^2+\frac{1}{2}
 \lambda_R \phi^2} \left[
1-\frac{\Lambda^3}{(\Lambda^2+m_R^2+\frac{1}{2} \lambda_R \phi^2)^{3/2}}
\right]
\label{zren}
\end{equation}

The flow with the coarse graining scale of the effective potential 
$V_{\Lambda}$ and of the wave function factor
$Z_{\Lambda}$ in the one loop approximation 
follows immediately from Eq.(\ref{V1loop}) and Eq.(\ref{Z1loop}),
\begin{eqnarray}
\Lambda {d V_{\Lambda}\over d \Lambda}&=&- {\Lambda^3 \over 4\pi^2}
\sqrt{\Lambda^2+V''}\nonumber \\
\Lambda {d Z_{\Lambda}\over d \Lambda} &=&-{\Lambda^3\over 64\pi^2}
{V'''^2\over (\Lambda^2+V'')^{5/2}}
\label {1loopdiffeq}
\end{eqnarray}
The equation for the effective potential has been previously 
obtained in
Ref. \cite{strick} using a blocking procedure. 

The study of the potential $V_{\Lambda}$ shows that it is possible to
have a non-trivial domain structure even in the symmetrical
phase of the theory ($m_R^2>0$) \cite{strick}. Indeed, for some range 
of the parameters
of the theory, it may happen that, although $m_R^2>0$, the squared
bare
mass is negative. In this case, the potential has a 
double well 
structure with two symmetrical non-zero minima for scales 
$\Lambda$ greater than a critical one $\Lambda_{cr}$, and has a
unique minimum at $\phi=0$ for smaller values of $\Lambda$
(see Fig. 1). The interpretation  of this fact 
is that the average field fluctuates around zero for scales 
$\zeta> \Lambda_{cr}^{-1}$ or 
around the non-zero minima for scales $\zeta<\Lambda_{cr}^{-1}$. 
The symmetrical phase therefore contains domains of size $\zeta \approx 
\Lambda_{cr}^{-1}$.
We remark that this phenomenon takes place in the symmetrical 
phase of the theory, and 
should not be confused with Spontaneous Symmetry Breaking
(SSB).

On the other hand, when SSB takes place ($m_R^2<0$), both the renormalized
one loop potential and wave function renormalization develop an
imaginary part for $\Lambda < \Lambda_{ssb}\equiv \sqrt 
{-m_R^2-\frac {1}{2}\lambda_R\phi^2}$. 
These imaginary parts generate non-real terms in the equations
of motion, and are artifacts of the adiabatic approximation.
They have nothing to do with the noise terms in the CTP
effective action that we have mentioned before.
It is worth noting that the wave function renormalization
diverges as
$\Lambda$ approaches $\Lambda_{ssb}$ from above. This clearly shows that 
the adiabatic approximation is not adequate to describe the
temporal evolution of the order parameter 
neither in the vicinity of $\Lambda_{ssb}$ nor for
$\Lambda <\Lambda_{ssb}$.

As pointed out in Ref.\cite{ww}, the imaginary part in the effective 
potential is a signal for instabilities towards the formation of
domains  of size
at least as great as  $\sqrt{-m_R^2}$. This issue has been addressed using
the Hartree approximation in Ref.\cite{hartree}
where it was shown that the size of
the domains for very weakly coupled theories can be much larger than 
the zero temperature
correlation length $\sqrt{-m_R^2}$.


\section{THE EXACT CTP RENORMALIZATION GROUP EQUATION}

In this Section we shall derive an exact (non-perturbative) 
renormalization group equation for the 
flow of the CGEA. The approach 
follows that of Wegner and Houghton for euclidean spacetime. We start by 
writing the CGEA for a scale $\Lambda - \delta \Lambda$, namely
\begin{equation}
e^{i S_{\Lambda - \delta \Lambda}(\phi_+,\phi_-)} \equiv \int 
\prod_{\Lambda_0>|{\vec q}| > \Lambda-\delta \Lambda}
 {\cal D}[\phi_+({\vec q},t)] 
{\cal D}[\phi_-({\vec q},t)] \; e^{i S_{cl}[\phi_+,\phi_-]}
\end{equation} 
The modes to be integrated can be splitted into two parts: one within the 
shell $\Lambda > |{\vec q}| > \Lambda - \delta \Lambda$ and another for 
modes with $\Lambda_0 > |{\vec q}| > \Lambda$. Expanding the action in 
powers of the modes within the shell, one has
\begin{eqnarray}
e^{i S_{\Lambda - \delta \Lambda}(\phi_+,\phi_-)} &=&
e^{i S_{\Lambda}(\phi_+,\phi_-)} \times \nonumber \\
&&\int \prod_{\Lambda >|{\vec q}| > \Lambda-\delta \Lambda} {\cal D}[\phi_+] 
{\cal D}[\phi_-] \; e^{i(S_1+S_2+ S_3)} 
\; e^{\frac{i}{2} \int^{'} \frac{d^3q}{(2\pi)^3} \int dt 
\frac{d}{dt} ( \phi_a(-{\vec q},t) \dot{\phi}_b({\vec q},t) g_{ab} )} 
\label{exp}
\end{eqnarray} 
where 
\begin{eqnarray}
S_1 &=& \int dt \int^{'} \frac{d^3q}{(2\pi)^3} \; \phi_a({\vec q},t) \;
\frac{ \partial S_{\Lambda} }{ \partial \phi_a (-{\vec q},t) } 
\nonumber \\
S_2 &=& \frac{1}{2} \int dt \; dt' \int^{'} \frac{d^3q}{(2\pi)^3} 
\; \phi_a({\vec q},t) \; \frac{\partial^2 S_{\Lambda}}
{\partial \phi_a(-{\vec q},t) \phi_b({\vec q},t')} \; 
\phi_b({\vec q},t') 
\end{eqnarray}
the prime in the momenta integrals meaning that integration is restricted to 
the shell. In the functional derivatives of $S_{\Lambda}$ 
(which contains modes whose wave vectors satisfy $|{\vec q}| < \Lambda$) 
the modes within the shell are set to zero. We use the notation
\begin{eqnarray}
\phi_a({\vec q},t) = \left( \begin{array}{c}
                \phi_+({\vec q},t) \\ \phi_-({\vec q},t)
                \end{array}
         \right) & \;  ; \;& 
g_{ab} = \left( \begin{array}{cc}
                1 & 0 \\ 0 & -1
                 \end{array} \right)
\end{eqnarray}

The $S_3$ term is cubic in the modes within the shell, and it can be proved 
that it does not contribute 
in the limit $\delta \Lambda \rightarrow 0$ (basically, 
this is because one is doing a one loop calculation for the shell modes). The 
functional integrals over the shell modes have the CTP boundary conditions. 
A comment about the last exponential factor 
in Eq.(\ref{exp}) is in order. Usually one discards 
it because it is a surface term, but in the CTP formalism it must be kept 
since the boundary conditions are that $\phi_+({\vec q},T) = 
\phi_-({\vec q},T)$ with $T \rightarrow \infty$ for the modes ${\vec q}$ 
within the shell. 

In order to evaluate the functional integrals we split 
the field as $\phi_a={\bar \phi}_a + \varphi_a$ and impose the boundary 
conditions on the ``classical" fields 
${\bar \phi}_{\pm}$, i.e. they vanish in the past $-T$ (negative and positive
frequencies respectively) and match in the Cauchy surface at time $T$.
The fluctuations $\varphi_a$ vanish both in the past and in the future. 
The classical fields are solutions to
\begin{equation}
(- \frac{d^2}{dt^2} - q^2) g_{ab} \bar{\phi}_b({\vec q},t) + 
\int dt' 
\frac{\partial^2 S_{int}}
{\partial \varphi_a(-{\vec q},t) \partial 
\varphi_b({\vec q},t')} \bar{\phi}_b({\vec q},t') = 0
\label{mod1}
\end{equation}
where we have splitted the CGEA as 
$S_{\Lambda}(\phi_{\pm}) = S_{kin}(\phi_{\pm})+S_{int}(\phi_{\pm})$ with
\begin{equation}
S_{kin} =  \int d^4x  \left[ \frac{1}{2}  (\partial_{\mu} \phi_+)^2 + 
\frac{i \epsilon}{2} \phi_+^2 \right] -
\int d^4x  \left[ \frac{1}{2} (\partial_{\mu} \phi_-)^2 -
\frac{i \epsilon}{2} \phi_-^2 \right]
\end{equation}
As before, in the 
functional derivatives the modes within the shell are set to zero.

Let $h_a$ be solutions to Eq.(\ref{mod1}), vanishing in the past and
satisfying an arbitrary normalization in the future, and let $\phi(\vec{q})$
be the common value the fields take in the future. We can then write
\begin{equation}
\bar{\phi}_a({\vec q},t) = \phi({\vec q}) 
\frac{h_a({\vec q},t)}{h_a({\vec q},T)}
\end{equation}
We first integrate over the common value $\phi(\vec{q})$ and then proceed with
the functional integration over the fluctuations $\varphi_a$ (both are
gaussian integrals with ``source'' terms). One finally gets

\begin{eqnarray}
\Lambda \frac{\partial S_{\Lambda}}{\partial \Lambda} &=& 
- \frac{i \Lambda}
{2 \delta \Lambda} \int^{'} \frac{d^3q}{(2\pi)^3} 
\ln \left( \frac{\dot{h}_{+}({\vec q},T)}{h_{+}({\vec q},T)} -
\frac{\dot{h}_{-}({\vec q},T)}
{h_{-}({\vec q},T)} \right) + \nonumber \\
&& \frac{\Lambda}{2 \delta \Lambda}  \int^{'} \frac{d^3q}{(2\pi)^3} 
\left( \frac{\dot{h}_{+}({\vec q},T)}{h_{+}({\vec q},T)} -
\frac{\dot{h}_{-}({\vec q},T)}
{h_{-}({\vec q},T)} \right)^{-1} \, 
\left( \int dt \frac{h_a(\vec{q},t)}{h_a(\vec{q},T)} 
\frac{\partial S_{\Lambda}}{\partial \varphi_a(-{\vec q},t)} \right)^{2} 
- \nonumber \\
&& \frac{i \Lambda}{2 \delta \Lambda} \ln det'(A_{ab}) + \nonumber \\
&& \frac{\Lambda}{2 \delta \Lambda} \int dt \; dt' \int^{'} \frac{d^3q}
{(2\pi)^3} \frac{\partial S_{\Lambda}}{\partial 
\varphi_a({\vec q},t)} A_{ab}^{-1}(-{\vec q},t;{\vec q},t') 
\frac{\partial S_{\Lambda}}{\partial \varphi_b({\vec q},t')} 
\label{exact}
\end{eqnarray}

The $2\times 2$ matrix $A_{ab}$ has the following elements
\begin{eqnarray}
A_{++}(-{\vec q},t;{\vec q\, '},t') &=& 
        (- \frac{d^2}{dt^2} - q^2 + i \epsilon) \delta(t-t') 
        \delta^3({\vec q}+{\vec q\,'}) + 
        \frac{\partial^2 S_{int}}{\partial \varphi_+(-{\vec q},t) 
        \partial \varphi_+({\vec q\,'},t')} \nonumber \\
A_{--}(-{\vec q},t;{\vec q\,'},t') &=& 
        (\frac{d^2}{dt^2} + q^2 + i \epsilon) \delta(t-t') 
        \delta^3({\vec q}+{\vec q\,'}) + 
        \frac{\partial^2 S_{int}}{\partial \varphi_-(-{\vec q},t) 
        \partial \varphi_-({\vec q\,'},t')} \nonumber \\
A_{+-}(-{\vec q},t;{\vec q\,'},t')&=& A_{-+}({\vec q\,'},t';{-\vec q},t)=
 \frac{\partial^2 S_{int}}{\partial \varphi_+(-{\vec q},t) 
        \partial \varphi_-({\vec q\,'},t')}
\end{eqnarray}
The primed determinant must be calculated as the 
product of the eigenvalues of $A_{ab}$ in 
a space of functions with wave vectors within the 
shell ($\Lambda -\delta \Lambda < |{\vec q}| < \Lambda$) 
and satisfying null conditions both in the past 
and in the future. Similar conditions are to be used to evaluate the 
inverse $A_{ab}^{-1}$.  

The equation (\ref{exact}) is exact in the sense 
that no perturbative approximation has 
so far been used. It is the main result of the paper. It is similar 
to its euclidean counterpart \cite{wegner}, but involves 
two fields and CTP boundary conditions. It contains {\it all} the 
information of the influence of the short wavelength modes on the long 
wavelength ones, and should be the starting point for a non-perturbative 
analysis of decoherence, dissipation, domain formation and out of 
equilibrium evolution. 
 

\section{DERIVATIVE EXPANSION}

The overwhelming complexity of the exact renormalization group 
equation means that 
in practice one is compelled to use some sort of truncation. The usual 
ones are expansions in the number of powers of the fields 
(see Ref. \cite{morris2} for a detailed analysis) or in derivatives 
of them \cite{morris3,enrique,tetradis}. In the following we 
shall make use of the derivative 
expansion approach.

We will prove that, within this approach, the 
exact RG Eq.(\ref{exact}) admits a solution of the form
\begin{equation}
S_{\Lambda}(\phi_+,\phi_-) = S_{\Lambda}(\phi_+) - 
S_{\Lambda}(\phi_-)
\label{ans}
\end{equation} 
Clearly this is not the most general form that can be 
imagined for the coarse grained action because contributions 
involving mixing of both fields are not taken into account. 
The main drawback of this approach is therefore that, as mentioned
in Section II,  we miss the stochastic 
aspects of the theory, such as dissipation and noise. However, 
the proposed form for the CGEA will be enough for 
studying the renormalization group flow of real time field theories. 

The great technical advantage of the  form Eq. (\ref{ans}) 
is that the second 
functional 
derivative of the action has no crossed terms, leading to a diagonal 
matrix $A_{ab}$ 
whose determinant is easily computed as the product of two determinants,
one for $A_{++}$ and one for $A_{--}$. Following Ref. \cite{cole} one 
can express both $det' A_{++}$ and $det' A_{--}$ as the 
product over momenta of a constant (momenta independent) times the
mode $h({\vec q},T)$ evaluated at the final time $T$. Therefore the last 
term of the exact RG equation can be written as
\begin{equation}
\ln det'(A_{ab}) = \ln [det'(A_{++}) det'(A_{--})] = 
\int^{'} \frac{d^3q}{(2\pi)^3} \ln( h_{+}({\vec q},T) h_{-}({\vec q},T) )
\end{equation}
The first and the third terms can then be casted in the form of a 
single logarithm, and we arrive at
\begin{eqnarray}
\Lambda \frac{\partial S_{\Lambda}}{\partial \Lambda} &=& 
- \frac{i \Lambda} {2 \delta \Lambda} \int^{'} \frac{d^3q}{(2\pi)^3} 
\ln({h_{-}({\vec q},T) \dot{h}_{+}({\vec q},T) - h_{+}}({\vec q},T) 
\dot{h}_{-}({\vec q},T)) 
\nonumber + \\
&&  \frac{\Lambda}{2 \delta \Lambda}  \int^{'} \frac{d^3q}{(2\pi)^3} 
\left( \frac{\dot{h}_{+}({\vec q},T)}{h_{+}({\vec q},T)} -
\frac{\dot{h}_{-}({\vec q},T)}
{h_{-}({\vec q},T)} \right)^{-1} \, 
\left( \int dt \frac{h_a(\vec{q},t)}{h_a(\vec{q},T)} 
\frac{\partial S_{\Lambda}}{\partial \varphi_a(-{\vec q},t)} \right)^{2} 
+ \nonumber \\
&&  \frac{\Lambda}{2 \delta \Lambda} \int dt \; dt' \int^{'} \frac{d^3q}
{(2\pi)^3} \frac{\partial S_{\Lambda}}{\partial 
\varphi_a({\vec q},t)} A_{ab}^{-1}(-{\vec q},t;{\vec q},t') 
\frac{\partial S_{\Lambda}}{\partial \varphi_b({\vec q},t')} 
\label{Slambda}
\end{eqnarray}

Note that the equations for the two modes $h_+$ and $h_-$ (Eq. (\ref{mod1})) 
simplify considerably, since the two equations are 
decoupled (a side point: it is easy to prove the one 
loop result Eq. (\ref{1looP})
starting from the above Eq. (\ref{Slambda})).
What we still have to prove is that the proposed form for the 
action makes the r.h.s. of the exact RG equation split in the same form. 

Next we make a derivative expansion of the interaction term. As our 
coarse graining explicitly breaks Lorentz invariance, 
we allow different coefficients for the 
temporal and spatial derivatives, namely
\begin{equation}
S_{int}(\phi_{\pm}) = \int d^4x [ - V_{\Lambda}(\phi_{\pm}) + 
\frac{1}{2} Z_{\Lambda}(\phi_{\pm})  \dot{\phi}^2_{\pm} -
\frac{1}{2} Y_{\Lambda}(\phi_{\pm})( {\vec \nabla} \phi_{\pm})^2 + \ldots ]
\end{equation}
We expand the fields around a time dependent background: $\phi_{\pm} = 
\phi_{\pm}(t) + \varphi_{\pm}({\vec x},t)$ and Fourier transform in 
space. 
We shall solve the 
Eq. (\ref{mod1}) for the modes to zeroth order in the fluctuations, i.e. 
we equate terms 
in the equations for $h_{\pm}$ that are
independent of $\varphi_{\pm}$'s.
Since the first functional derivative of the CGEA (S') is 
linear in the fluctuations $\varphi_{\pm}$, we 
put $S'=0$ and keep the $\varphi_{\pm}$-independent contributions to 
$S_{int}''$. After some little algebra and functional derivations, we get
\begin{eqnarray}
\frac{\partial^2 S_{int}}{\partial \varphi({\vec q},t) \partial 
\varphi(-{\vec q}',t')}
&=& [ -V'' - \frac{1}{2} \dot{\phi}^2 Z'' - Y q^2 -  Z' \dot{\phi}
\frac{d}{dt} - Z \frac{d^2}{dt^2} - \ddot{\phi} Z' + \ldots ] 
\times \nonumber \\
&&\delta(t-t') \delta^3({\vec q}-{\vec q}')
\end{eqnarray}
where the primes denote derivation with respect to the field and the 
ellipsis denote terms linear in the fluctuations. 
In this expression and hereafter we omit (unless explicitely stated) 
the ${\pm}$ subscripts in the background 
fields $\phi_{\pm}(t)$, in the 
potential $V_{\Lambda}(\phi_{\pm}(t))$, and in the wave 
function factors $Z_{\Lambda}(\phi_{\pm}(t))$ and 
$Y_{\Lambda}(\phi_{\pm}(t))$. Note that  
the effective mass of the modes depends on the time-dependent background
$\phi(t)$.
The equations of motion for the modes $h_a$ become localized and take the 
form of
harmonic oscillators with variable frequency and a damping term.  
The boundary conditions to be imposed are the aforementioned CTP ones. 

If one defines new modes as
$f({\vec q},t) = (1+Z_{\Lambda})^{1/2} h({\vec q},t)$, the damping 
terms cancel out and the new modes are harmonic oscillators with frequency 
\begin{equation}
w_{q}^2(t)= q^2 \, \frac{1+Y_{\Lambda}}{1+Z_{\Lambda}} + 
\frac{V_{\Lambda}''}{1+Z_{\Lambda}} + 
\frac{1}{4} \frac{Z_{\Lambda}'^2}{(1+Z_{\Lambda})^{2}} \, \dot{\phi}^2 +
\frac{1}{2} \frac{Z_{\Lambda}'}{1+Z_{\Lambda}} \,  \ddot{\phi}
\end{equation}
Using an adiabatic expansion for the modes,
\begin{equation}
h_{\pm}({\vec q},t) = (1+Z_{\Lambda})^{-1/2} 
\frac{1}{\sqrt{2 W_{\pm}({\vec q},t)}} e^{\pm i 
\int_{-T}^{t} W_{\pm}({\vec q},t') dt'}
\end{equation}
we can evaluate the logarithmic term in the r.h.s. of the exact RGE
(\ref{Slambda}). 
The other terms are quadratic in the 
fluctuations and do not contribute to the order we are working. 
We have
\begin{eqnarray}
&h_{-}({\vec q},T) \dot{h}_{+}({\vec q},T) - h_{+}({\vec q},T) 
\dot{h}_{-}({\vec q},T) =& \nonumber \\
&e^{i \int_{-T}^{T} [W_{+}({\vec q},t)-W_{-}({\vec q},t)] dt} \times 
\left\{
\frac{\left[
        -\frac{Z_+'}{2(1+Z_+)} \dot{\phi}_+ - \frac{\dot{W}_+}{2 W_+} 
+ i W_+
      \right]-
      \left[
        -\frac{Z_-'}{2(1+Z_-)} \dot{\phi}_- - \frac{\dot{W}_-}{2 W_-}
- i W_-
      \right] }
     {2 \sqrt{W_+ W_- (1+Z_+) (1+Z_-)} }
     \right\}_{t=T}
\end{eqnarray}
Note that, as in the one loop case, the plus 
and minus fields do decouple, up to a factor evaluated at $t=T$. This 
factor is just a surface contribution which, upon taking logarithm, 
is irrelevant for the equations of motion. On the contrary, the first factor 
does depend on the whole history of the fields and it is consistent with the 
proposal for the CGEA Eq.(\ref{ans}). Hence the RGE can be casted in the form 
\begin{eqnarray}
&&\Lambda \int dt 
\left\{
\left[
- \frac{d V_{\Lambda}(\phi_+)}{d \Lambda} + \frac{1}{2} 
\frac{d Z_{\Lambda}(\phi_+)}{d \Lambda} \dot{\phi}_+^2 
\right]
-
\left[
- \frac{d V_{\Lambda}(\phi_-)}{d \Lambda} + \frac{1}{2} 
\frac{d Z_{\Lambda}(\phi_-)}{d \Lambda} \dot{\phi}_-^2 
\right]
\right\} = \nonumber \\
&& \frac{ \Lambda}{2 \delta
\Lambda} \int^{'} \frac{d^3q}{(2\pi)^3}  \int 
[W_{+}({\vec q},t)-W_{-}
({\vec q},t)]dt
\end{eqnarray}

In the adiabatic expansion, the $W$'s read
\begin{equation}
W^2=A_{\Lambda} + B_{\Lambda}
\dot{\phi}^2(t)+ C_{\Lambda} \ddot{\phi}(t)
\end{equation} 
where the coefficients are 
\begin{eqnarray}
A_{\Lambda} &=& \Lambda^2 \frac{1+Y_{\Lambda}}{1+Z_{\Lambda}} +
     \frac{V_{\Lambda}''}{1+Z_{\Lambda}} \nonumber \\
B_{\Lambda} &=& \frac{Z_{\Lambda}'^2}{4(1+Z_{\Lambda})^2} + 
     \frac{5 A_{\Lambda}'^2}{16 A_{\Lambda}^2} - \frac{A_{\Lambda}''}
{4 A_{\Lambda}} \nonumber \\
C_{\Lambda} &=& \frac{Z_{\Lambda}'}{2(1+Z_{\Lambda})} 
- \frac{A_{\Lambda}'}{4 A_{\Lambda}} 
\end{eqnarray}
Integrating by parts we get 
\begin{equation}
\int dt \left\{
- \Lambda \frac{dV_{\Lambda}}{d\Lambda} + 
\frac{1}{2} \Lambda \frac{dZ_{\Lambda}}{d\Lambda} {\dot \phi}^2 \right\}
=\frac{\Lambda^3}{4 \pi^2} \int dt 
\left\{
\sqrt{A_{\Lambda}} +
\frac{1}{2} \dot{\phi}^2 
\left[
\frac{B_{\Lambda}}{\sqrt{A_{\Lambda}}} - 
\left( \frac{C_{\Lambda}}{\sqrt{A_{\Lambda}}} \right)'
\right]
\right\}
\end{equation}
Therefore the dependence of the potential and the wave function
renormalization on the infrared scale is given by
\begin{eqnarray}
\Lambda \frac{dV_{\Lambda}}{d\Lambda} &=& -\frac{\Lambda^3}{4 \pi^2} 
\sqrt{\Lambda^2 \frac{1+Y_{\Lambda}}
{1+Z_{\Lambda}} + \frac{V_{\Lambda}''}{1+Z_{\Lambda}}} \nonumber \\
\Lambda \frac{dZ_{\Lambda}}{d\Lambda} &=& \frac{\Lambda^3}{4 \pi^2} 
\left[
\frac{B_{\Lambda}}{\sqrt{A_{\Lambda}}} - 
\left( \frac{C_{\Lambda}}{\sqrt{A_{\Lambda}}} \right)'
\right]
\label{flowing}
\end{eqnarray}
Remember that these equations are valid both for the $\phi_+$ field 
and for the $\phi_-$ field.

The above equations are the main result of this section. They describe the flow
of the coarse grained action with the infrared scale in the derivative 
expansion 
of the exact CTP renormalization group equation. It is
interesting to note that the higher derivative terms
modify the differential equation for the effective
potential. 

We have obtained two equations for the three independent 
unknown functions $V_{\Lambda}$, $Z_{\Lambda}$ and $Y_{\Lambda}$. 
In order to find an additional relation between the 
spatial and temporal wave funtion renormalization 
functions $Z_{\Lambda}$ and $Y_{\Lambda}$, it is 
necessary to write the exact RGE up to quadratic order in the 
fluctuations. We will not present this long calculation here. For simplicity,
we will assume that $Z_{\Lambda}$ and $Y_{\Lambda}$ are small numbers, 
and therefore we will set them to zero on the r.h.s. of Eq.(\ref{flowing}). 
This assumption will be confirmed by the numerical calculations 
performed in the symmetric phase of the theory. Note that in this 
approximation we recover the RG improved equation proposed in 
Ref. \cite{strick} for the coarse grained effective potential. 

There are other  points which are worth noting. The first
one is that, when we substitute $V_{\Lambda}$, $Z_{\Lambda}$
and $Y_{\Lambda}$ by their classical values 
$V_{\Lambda}=V \, , \, Z_{\Lambda}=Y_{\Lambda}=0$, on the
r.h.s. of both equations we obtain the one loop
evolution equations (Eq. (\ref{1loopdiffeq})). The second comment is
about renormalization. In the one loop calculation
it was possible to take the limit $\Lambda_{0}\rightarrow\infty$.
The infinities were absorbed into the bare mass and coupling constant.
There was no need for wave function renormalization. However,
in this non-perturbative calculation it is not possible to 
renormalize the theory
(as is the case for Hartree, Gaussian and $1/N$ approximations).
For these reasons we keep $\Lambda_{0}$ as a large
(compared with the mass) but finite number. 

As illustration we consider the $\lambda \phi^4$ field theory. 
The differential equations must be solved with
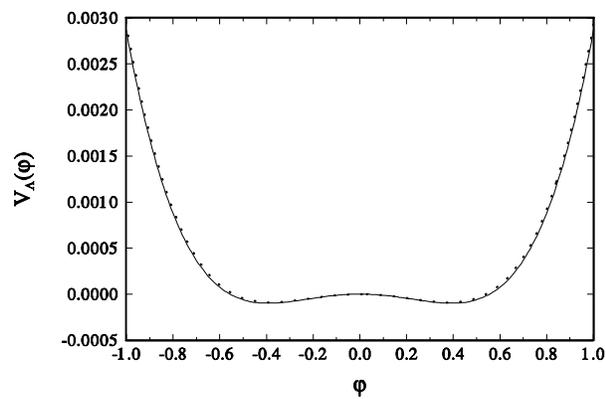
the classical initial conditions $V_{\Lambda_{0}}=V$, $Z_{\Lambda_0}=0$
and $Y_{\Lambda_{0}}=0$. To this end we have written a simple code which 
evolves the equations from the UV scale down to the desired IR scale. 
We have plotted the results in Figs. 1 and 2, where we show the 
difference between
the one loop and the RG improved solution for the effective potential
and the wave function renormalization. Note that, at least 
in the symmetric phase of the theory, the results are 
consistent with the assumption
$Z_{\Lambda} \ll 1$. 

Once the functions $V_{\Lambda}$ and $Z_{\Lambda}$ are known,
one can write the effective dynamical equations for the order parameter.
These equations will be valid as long as $\phi$ is slowly varying 
and $Z_{\Lambda} \ll 1$.Therefore, as in the one loop approximation, 
the derivative expansion may be inadequate in the case of SSB for scales
in the vicinity of and lower than $\Lambda_{ssb}$.

\section{CONCLUSIONS}

The CTP coarse grained effective action contains all the information
about the influence of the short wavelength sector on
the long wavelength sector of the theory. In principle it is
possible to 
derive  from it a Langevin equation for the order parameter. This 
equation can be used to analyze domain formation and growth
and, in general, the non-equilibrium aspects of phase transitions.

In this paper we have derived an exact evolution equation for
the dependence of the CGEA with the coarse graining scale. This
renormalization group equation (Eq. (\ref{exact})) is our main result. 
We expect this equation
to be a useful tool to generate non-perturbative approximations
for the CGEA.

In order to show  a simple application of the exact CTP renormalization 
group equation,
we have solved it using a derivative expansion. 
We have obtained a RG improvement to the effective potential
$V_{\Lambda}(\phi)$, that coincides with the one proposed in Ref. \cite{strick},
and an improvement to the one loop wave function renormalization $Z_{\Lambda}$.
   
Within the derivative expansion
approach, the CGEA is of the form $S_{\Lambda}(\phi_+,\phi_-) 
= S_{\Lambda}(\phi_+) - 
S_{\Lambda}(\phi_-)$, i.e., it does not contain 
neither dissipative nor noise terms. Only diffussive effects
are included. As far as a calculation beyond the adiabatic 
approximation is concerned, 
we expect that, as soon as we decrease the scale from $\Lambda_{0}$,
dissipative and noise terms will grow up: the CGEA
will develop an imaginary part (related to noise)
and a real part containing interactions between
the $\phi_{\pm}$ fields (dissipation).
This can be easily checked both in the one loop approximation and from the
exact RGE. Indeed, Eq.(\ref{1looP}) shows that the one loop CGEA is in 
general non-real unless one uses the adiabatic approximation. On the
other hand, the real and imaginary parts of the CGEA are not decoupled
in Eq. (\ref{exact}), and a non-vanishing real part at $\Lambda=\Lambda_0$
will induce an imaginary part at lower scales. 

We are currently investigating these issues. In particular,
we are interested in finding a non-perturbative, $\Lambda$-dependent
fluctuation-dissipation relation from the exact RGE. 
Extensions to the cases of finite temperature and curved spaces are
also under investigation.

\section{Acknowledgements}

We would like to thank B.L. Hu, F. C. Lombardo, J.P. Paz
and F. M. Spedalieri for useful discussions. F.D.M. enjoyed
the hospitality of the General Relativity Group
at the University of Maryland, where part of this work
was done. This work was supported by Universidad
de Buenos Aires, CONICET and Fundaci\' on Antorchas.

\newpage

\hskip5cm FIGURE CAPTIONS

\bigskip

FIG. 1: The coarse grained effective potential $V_{\Lambda}$ for 
$\Lambda_0=10$, $m_R^2=10^{-4}$ and $\lambda_R=0.1$. 
Solid and dotted lines are the RG improved
and one loop results respectively.

FIG. 2: The coarse grained wave funtion renormalization $Z_{\Lambda}$ for 
$\Lambda_0=10$, $m_R^2=10^{-4}$ and $\lambda_R=0.1$. Solid and 
dotted lines are the RG improved
and one loop results respectively.

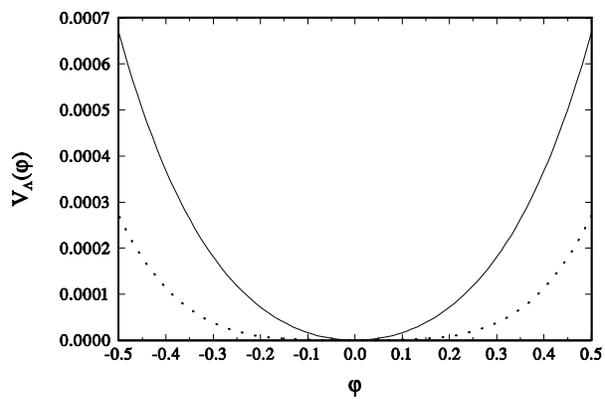
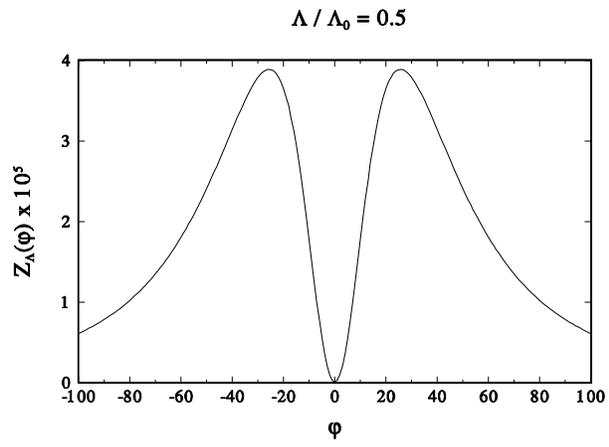
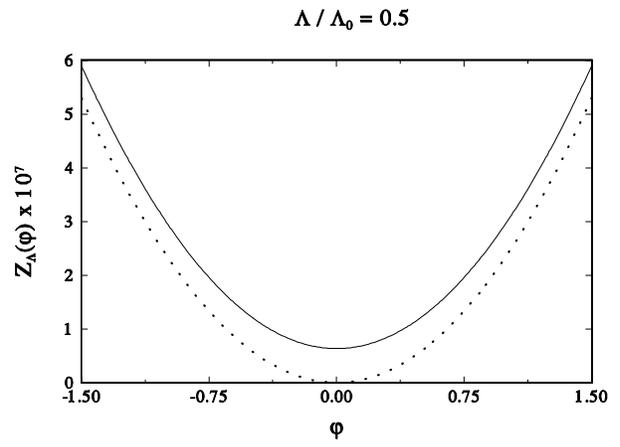
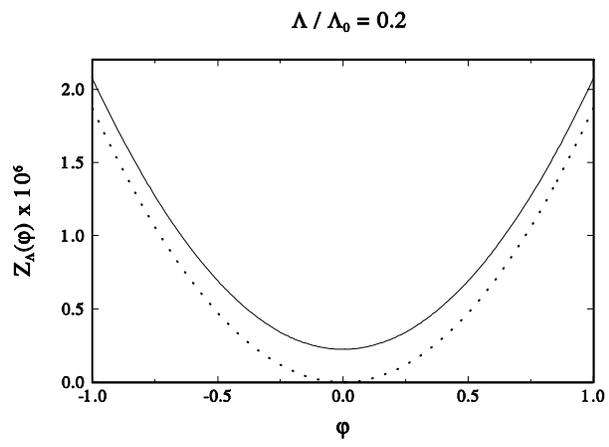
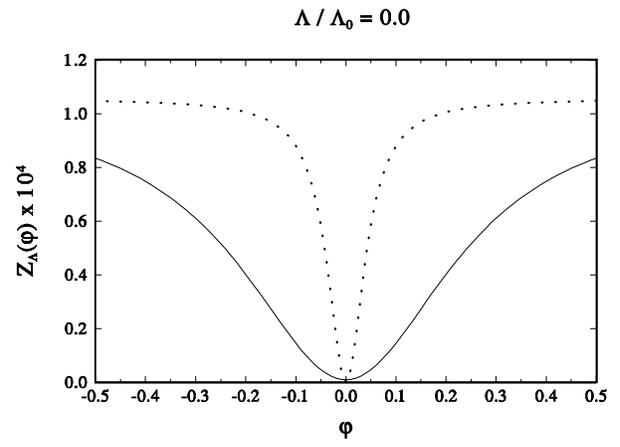
\end{document}